\documentclass[preprint,12pt]{elsarticle}
\usepackage{amsmath}
\usepackage{graphicx}  
\usepackage{epstopdf}   
\usepackage{lineno,hyperref}
\modulolinenumbers[5]
\usepackage{color}

\begin{document}

\begin{frontmatter}

\title{Non-isothermal stress relaxation in conventional and high-entropy metallic glasses and its relationship to the mixing and excess entropy}


\author[VSPU]{G.V. Afonin\corref{cor1}}
\cortext[cor1]{{Corresponding author}} 
\ead{afoningv@gmail.com, Tel/fax:+7-473-255-24-11} 
\author[VSPU]{S.L. Scherbakov}
\author[VSPU]{R.A. Konchakov}
\author[IFTT]{N.P. Kobelev}
\author[NPU]{J.B. Cui}  
\author[NPU]{J.C. Qiao} 
\author[VSPU]{V.A. Khonik}

\address[VSPU] {Department of General Physics, State Pedagogical
University,  Lenin St. 86, Voronezh 394043, Russia}
\address[IFTT] {Institute of Solid State Physics RAS, Chernogolovka 142432, Russia}
\address[NPU] {School of Mechanics, Civil Engineering and Architecture, Northwestern Polytechnical University, Xi’an, 710072, China}

\begin{abstract}
We performed  calorimetric and  torsion stress relaxation measurements upon linear heating of six conventional and high-entropy metallic glasses with the mixing entropy $\Delta S_{mix}$ ranging from $0.86R$ to $1.79R$ ($R$ is the universal gas constant). It is shown that high-entropy metallic glasses ($\Delta S_{mix}> 1.5\;R$) exhibit significantly greater resistance to stress relaxation. Based on calorimetric data, we calculated the excess entropy of glass relative to the counterpart crystalline state and introduced an entropy-based dimensionless parameter $\Delta_S$, which  characterizes the rise of the entropy and structural disordering of glass in the supercooled liquid region. It is shown that the depth of stress relaxation at a given temperature decreases with $\Delta S_{mix}$ but increases with $\Delta_S$. Possible reasons for this relationship are discussed.  
\end{abstract}

\begin{keyword}
\texttt{metallic glasses, calorimetry, stress relaxation, mixing entropy, excess entropy}
\end{keyword}

\end{frontmatter}

\section{Introduction}

High-entropy alloys were first developed in the early 2000s and continue to attract researchers' attention due to their unique properties compared to conventional structural alloys, i.e. high toughness and tensile ductility, ultra-high microhardness, exceptional wear resistance, corrosion and oxidation resistance, heat resistance, superelasticity, etc  \citep{Hofmann2008,Zhao2024,Tsaia2014, Riechers2025, Yaacoub2020}. High-entropy alloys are usually characterized by the mixing entropy  $\Delta S_{mix}=-R\sum^N_{i=1}{c_ilnc_i}$ , where $R$ is the universal gas constant, $c_i\ $  is the atomic fraction of the $i-$th element, and $N$ is the number of elements. The mixing entropy $\Delta S_{mix}$ thus defined characterizes  how much disorder is introduced upon alloy preparation. Alloys are usually considered to be high-entropy provided that $\Delta S_{mix}\ge 1.5\ R$, which is achieved in the compounds containing at least five main metallic components, each with a fraction of 5 to 35 at.$\%$ \cite{Yeh2004,GeorgeNatRevMater2019}. The maximum of the mixing entropy is reached in the equiatomic compositions. 

In the early 2010s is was shown that rapid quenching of high-entropy melts can lead to the formation  of a non-crystalline state \cite{Wang2014}. These non-crystalline alloys have been referred to as high-entropy metallic glasses (HEMGs) \cite{Wang2014}. Numerous studies over the past decade  showed that HEMGs can combine a number of unique properties common to both conventional metallic glasses and crystalline high-entropy alloys: extremely high elastic limit, strength and hardness,  wear/scratch and corrosion resistance, superplastic formability  as well as the ability for polishing to an almost atomically smooth surface, etc. \cite{Tong2016, Chen2021}.

The non-crystalline structure of HEMGs provides a thermodynamic driving force for structural relaxation, which leads to changes in various physical properties during physical aging. However, despite intensive studies over the past decade \cite{Luan2023}, the role of the high-entropy state in the relaxation behavior of HEMGs and their physical properties  still remains uncertain \cite{Afonin2024}. In particular, it is largely unclear whether and how the mixing entropy $\Delta S_{mix}$ affects HEMGs' relaxation ability  and are there any specific parameters suitable for a characterization of  physical properties of HEMGs. The current research aims to address these issues.
We studied stress relaxation behavior of metallic glasses upon systematically increasing $\Delta S_{mix}$  and on the basis of calorimetric data introduced a new entropy-based parameter $\Delta_S$, which determines, together with $\Delta S_{mix}$, their relaxation ability.  We found that the depth of stress relaxation at given normalized temperature systematically decreases with $\Delta S_{mix}$ but increases with $\Delta_S$. Possible reasons for this behavior are discussed.

\section{Experimental}

\begin{table}[t]
\begin{center}
\footnotesize
\caption{Parameters of MGs under investigation: mixing entropy $\Delta S_{mix}/R $, excess entropy $\Delta S_{scl}$,  excess entropy  $\Delta S_{T_{g}}$ and entropy rise in the SCL range $\Delta_S = (\Delta S_{scl}-\Delta S_{Tg})/R$. } 
\begin{tabular}{p{5pt}p{169pt}ccccp{25pt}}

\hline
\hline
 N& Glass composition & $\Delta S_{mix}/R $  & $\Delta S_{scl}$ & $\Delta S_{T_{g}}$ &   $\Delta_S$   \\
  & at.$\%$ &  & $[\frac{J}{K\times mol}]$& $[\frac{J}{K\times mol}]$ &  \\
\hline
\hline

1 & $Cu_{50}Zr_{45}Al_{5}$   & 0.86 & 5.6 & 4.7 & 0.11  \\
2 & $Zr{}_{46}Cu_{36.8}Ag_{9.2}Al_{8}$ & 1.15 & 7.0 & 5.7 & 0.16   \\
3 & $Pd_{43.2}Cu_{28}Ni_{8.8}P_{20}$ & 1.25 & 6.3 & 4.5 & 0.22  \\
4 & $Zr_{35}Hf_{13}Al_{11}Ag_{8}Ni_{8}Cu_{25}$ & 1.63 &  3.5 &  2.9 & 0.072  \\
5 & $Zr_{40}Hf_{10}Ti_{4}Y_{1}Al_{10}Cu_{25}Ni_{7}Co_{2}Fe_{1}$ & 1.66 & 3.4 & 2.9 & 0.060   \\
6 & $Zr_{16.67}Hf_{16.67}Al_{16.67}Co_{16.67}Ni_{16.67}Cu_{16.67}$ & 1.79 & 4.8 & 4.7  & 0.012  \\
\hline
\end{tabular}
\end{center}
\label{onecolumntable}
\end{table}

Six metallic glasses (MGs),  Cu$_{50}$Zr$_{45}$Al$_{5}$,  Zr$_{46}$Cu$_{36.8}$Ag$_{9.2}$Al$_{8}$, Pd$_{43.2}$Cu$_{28}$Ni$_{8.8}$P$_{20}$, Zr$_{35}$Hf$_{13}$Al$_{11}$Ag$_{8}$Ni$_{8}$Cu$_{25}$, Zr$_{40}$Hf$_{10}$Ti$_{4}$Y$_{1}$Al$_{10}$Cu$_{25}$Ni$_{7}$Co$_{2}$Fe$_{1}$ and 
Zr$_{16.67}$Hf$_{16.67}$Al$_{16.67}$Co$_{16.67}$Ni$_{16.67}$Cu$_{16.67}$ (at.\%) listed  in Table 1 were chosen for the investigation. The first three MGs constitute conventional glass compositions with the mixing entropy $\Delta S_{mix}/R$ increasing from  0.86 to 1.25 while three latter MGs are high-entropy with $1.63\leq \Delta S_{mix}/R\leq 1.79$. It is to be noted that this composition range covers almost complete $\Delta S_{mix}/R$-range available experimentally. All MGs were produced  by single-roller spinning as 25--35 $\mu$m ribbons and X-ray checked to be fully amorphous. 

Differential scanning calorimetry (DSC) was performed using a Hitachi DSC 7020 in flowing $N_2$ (99.999$\%$ pure) atmosphere. The instrument was calibrated using the melting points and enthalpies of 99.99$\%$ pure In, Sn, Pb and Al. Every glass composition was tested as follows: \textit{i}) an initial sample was heated with empty reference DSC cell  up to the temperature of the complete crystallization $T_{cr}$ and cooled down to room temperature; this sample was next moved into the reference cell; \textit{ii}) a new initial sample of nearly the same mass (50-70 mg) was tested up to $T_{cr}$. This procedure allowed calculating the differential heat flow $\Delta W=W_{gl}-W_{cr}$, where $W_{gl}$ and $W_{cr}$ are the heat flows coming from glass and maternal (counterpart) crystal, respectively. Then, one can calculate the excess entropy of glass using the relationship suggested recently \cite{MakarovJPCM2021,MakarovJLETT2022},

\begin{equation}
\Delta S(T)= \frac{1}{\dot{T}}\int_{T}^{T_{cr}}\frac{\Delta W(T)}{T}dT, \label{Scalc}
\end{equation}
where $\dot{T}$ is heating rate. It is to be mentioned that if temperature $T\rightarrow T_{cr}$ then $\Delta S\rightarrow 0$ and, therefore, $\Delta S$ thus determined constitutes the excess entropy of glass with respect to the counterpart crystal.

In this work, we chose stress relaxation as a method to test the relaxation ability of MGs characterized by different mixing entropy. The stress relaxation method is often used in experimental studies, but it is only done under isothermal conditions and in the tensile mode. For non-isothermal experiments (i.e. at a constant heating rate) in the tensile mode, this method is practically unsuitable because of  thermal expansion of the testing machine, which corrupts  measurement results making them completely unreliable. However, if measurements are taken in the torsional mode, thermal expansion has no effect on the stress relaxation data. In this case, it is possible to measure the relaxation kinetics in the whole temperature range, starting from room temperature up to $T_{cr}$ or even higher temperatures. However, this experimental technique is rather complicated and we know just a few examples of non-isothermal measurements of torsional stress relaxation  \cite{BobrovPhysSolState2004,NguenPSS2009,AfoninJNCS2012}.               
 
\begin{figure}[t]
\begin{center}
\includegraphics[scale=1.3]{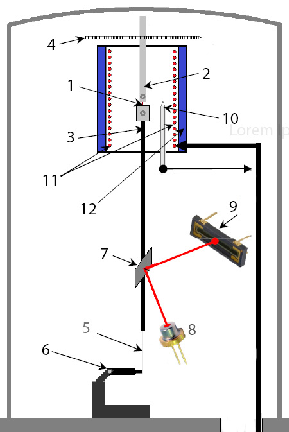}
\caption[*]{\label{Fig1.eps} Schematic diagram of the torsion testing micromachine. \textit{1} - sample, \textit{2},\textit{3} - non-deformable rods with microgrips, \textit{4} - torsion loading gear wheel, \textit{5} - quartz filament, \textit{6} - elastic plate, \textit{7} - mirror, \textit{8} - laser diode, \textit{9} - position sensitive diode, \textit{10} - thermocouple, \textit{11} - heater, \textit{12} - water cooling.   } 
\end{center}
\end{figure}

In this investigation, torque relaxation measurements were carried out using a laboratory-made torsion testing micromachine, a schematic diagram of which is shown in Fig.\ref{Fig1.eps}. A rectangular sample \textit{1} (gauge length 2--3 mm) was cut from glassy ribbon using special guillotine shears and  attached using special microgrips to two non-deformable rods \textit{2} and \textit{3}. The upper rod \textit{2} was connected with a gear wheel \textit{4}, which was used to produce torsion deformation. The lower rod \textit{3} was glued to the upper end of a 150 $\mu m$ calibrated quartz filament \textit{5}, which, in turn, was  glued to an elastic plate \textit{6}. This plate compensates for the thermal expansion of the system (rods + sample + quartz filament) and ensures that a very small tensile stress is applied to the sample. Loading by the gear wheel \textit{4} leads to the torsion deformation of the sample \textit{1} and quartz filament \textit{5} while the rods \textit{2} and \textit{3} remain non-deformable (due to their large cross-section). The light beam  sent by a laser diode \textit{8} reflects from a mirror \textit{7} and falls on a Hamamatsu S1352 position-sensitive diode (PSD) \textit{9}, which continuously records the displacement $x$ of the light spot. The torsion angle then becomes $\alpha \approx x/2r$, where $r$ is the distance between the mirror and PSD. The applied torque is  calculated as $M=\frac{\pi G_q\;d_q^4}{32\;l_q}\alpha$, where $d_q$ and $l_q$ are the diameter and length of the quartz filament and $G_q=31$ GPa is the shear modulus of quartz. The initial torque is determined as $M_0=\sigma abc^2$, where $b$ and $c$ are the long and short side of sample's cross-section (determined by an optical microscope), respectively,  $a=\left(3+1.8\frac{b}{c}\right)$ and $\sigma$ is the stress on sample's surface, which was accepted to be 250 MPa. 

Torque measurements were performed in a  vacuum of about 0.01 Pa at a heating rate of 3 K/min.  The measurements were started just after loading at room temperature. During the experiment, the total torsion angle was kept constant. Stress relaxation in the sample reduces the torque applied to the quartz filament, which in turn changes the displacement $x$ measured by the PSD. The relaxed sample was prepared by preliminary heating in the unloaded state to a temperature of $\approx 20$ K above the glass transition temperature $T_{g}$ and subsequent cooling back to room temperature. 

\section{Results}

\begin{figure}[t]
\begin{center}
\includegraphics[scale=0.8]{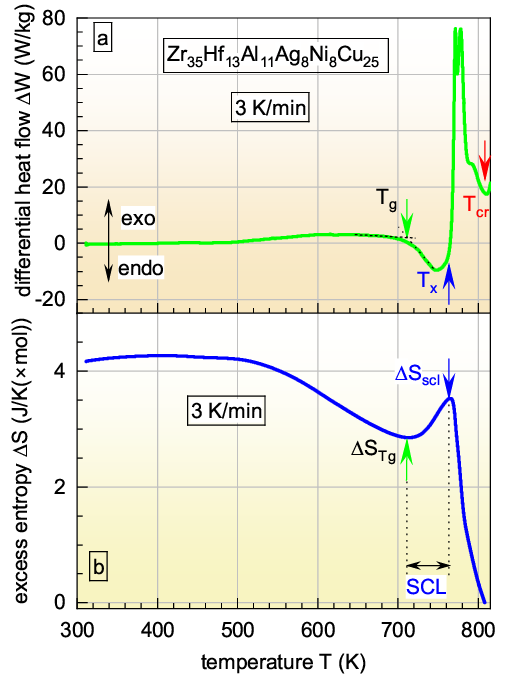}
\caption[*]{\label{Fig2.eps} (a) Differential heat flow of high-entropy glassy Zr$_{35}$Hf$_{13}$Al$_{11}$Ag$_{8}$Ni$_{8}$Cu$_{25}$ ($\Delta S_{mix}/R=1.63$). The characteristic  temperatures, glass transition $T_{g}$,  crystallization onset $T_{x}$ and complete crystallization $T_{cr}$, are indicated by the arrows. (b) Temperature dependence of the excess entropy $\Delta S$ calculated with Eq.(\ref{Scalc}) using the differential heat flow $\Delta W$ shown in panel (a). The characteristic entropies, $\Delta S_{T_{g}}$ and $\Delta S_{scl}$ as well as the SCL range are indicated.} 
\end{center}
\end{figure} 

Panel (a) in Fig.\ref{Fig2.eps} shows the differential heat flow $\Delta W(T)$ for high-entropy  glassy Zr$_{35}$Hf$_{13}$Al$_{11}$Ag$_{8}$Ni$_{8}$Cu$_{25}$, which is typical for the MGs in the present study. A small exothermal effect below the glass transition temperature $T_g$ (shown by the arrow) corresponds to structural relaxation while endothermal reaction in the range $T_g\leq T\leq T_x$ ($T_x$ is the crystallization onset temperature, indicated by the arrow) represents the supercooled liquid (termed "SCL" hereafter) state. Above $T_x$, a strong exothermal crystallization is observed, which terminates at a temperature $T_{cr}$. 

Panel (b) in Fig.\ref{Fig2.eps}  gives temperature dependence of the excess entropy $\Delta S$  calculated with Eq.(\ref{Scalc}) using $\Delta W$-data given in panel (a). It is seen that $\Delta S$ is nearly constant at temperatures below 500 K. At higher temperatures, $\Delta S$ decreases that corresponds to an increase of structural order due to exothermal relaxation captured by DSC as shown in panel (a). The excess entropy reaches a minimum at $T_g$, which we term as $\Delta S_{Tg}$ hereafter. Heating in the SCL range (i.e. at temperatures $T_g\leq T \leq T_x$) results in a rapid rise of the excess entropy, which reflects the increase of structural disorder in this range. This entropy at the temperature $T=T_{x}$ is designated as $\Delta S_{scl}$.

\begin{figure}[t]
\begin{center}
\includegraphics[scale=0.8]{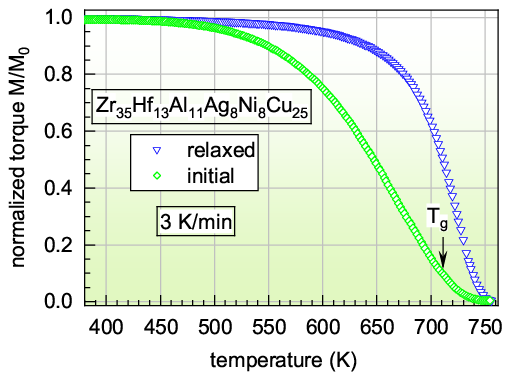}
\caption[*]{\label{Fig3.eps} Relaxation of the normalized torque $M/M_0$ for high-entropy $Zr_{35}Hf_{13}Al_{11}Ag_{8}Ni_{8}Cu_{25}$ glass in the initial and relaxed states. The calorimetric glass transition temperature $T_g$ is indicated by the arrow. It is seen that preliminary relaxation provides a strong shift of the relaxation towards higher temperatures.} 
\end{center}
\end{figure}

Figure \ref{Fig3.eps} gives temperature dependences of the normalized torque $M/M_0$ ($M_0$ is the initial torque applied at room temperature) for the same glass.  It is seen that there is almost no torque relaxation upon heating to 500 K, which corresponds to the onset of exothermal relaxation (see Fig.\ref{Fig2.eps}a). Heating to higher temperatures results in a significant relaxation, such that the relaxation depth $\Delta_M=1-M/M_0$ reaches approximately 90\% near the glass transition temperature $T_g$. Preliminary annealing by heating into the SCL range and cooling back to room temperature results in a significant shift of the relaxation curve towards high temperatures. Quite similar data were obtained for other MGs under investigation. 

\begin{figure}[t]
\begin{center}
\includegraphics[scale=0.8]{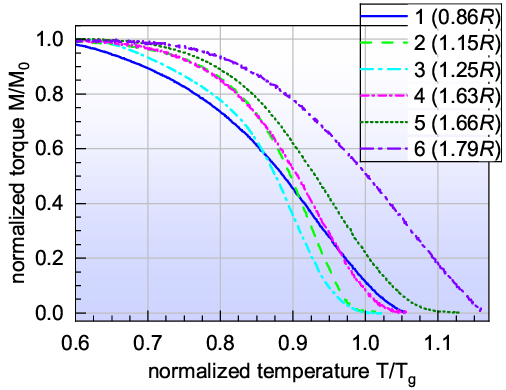}
\caption[*]{\label{Fig4.eps} Relaxation kinetics of the normalized torque $M/M_0$ as a function of the normalized temperature $T/T_g$. Glass compositions are given by the numbers according to Table 1.  The corresponding mixing entropies $S_{mix}$ are indicated in parentheses.} 
\end{center}
\end{figure} 

\section{Discussion}

In order to compare the relaxation kinetics for different MGs, it is convenient to normalize the temperature by the glass transition temperature $T_g$. This is done in Fig.\ref{Fig4.eps}, which presents the normalized torque $M/M_0$ as a function of the normalized temperature $T/T_g$ for all MGs under investigation. Glass compositions are indicated by the numbers according to Table 1 while the corresponding mixing entropies $\Delta S_{mix}$ are given in parenthesis. It is seen that, for instance, the normalized torque in glass 1 ($\Delta S_{mix}=0.86R$) decreases to $M/M_0\approx 0.45$ at a temperature $T/T_g\approx 0.9$. On the other hand, the high-entropy glass 6 displays the same fall of the normalized torque at significantly higher temperature, $T\approx 1.05\;T_g$. The general trend of the data given in Fig.\ref{Fig4.eps}, therefore, is quite clear: an increase of the mixing entropy $\Delta S_{mix}$ leads to a significant shift of the relaxation curves towards higher temperatures.     

\begin{figure}[t]
\begin{center}
\includegraphics[scale=0.45]{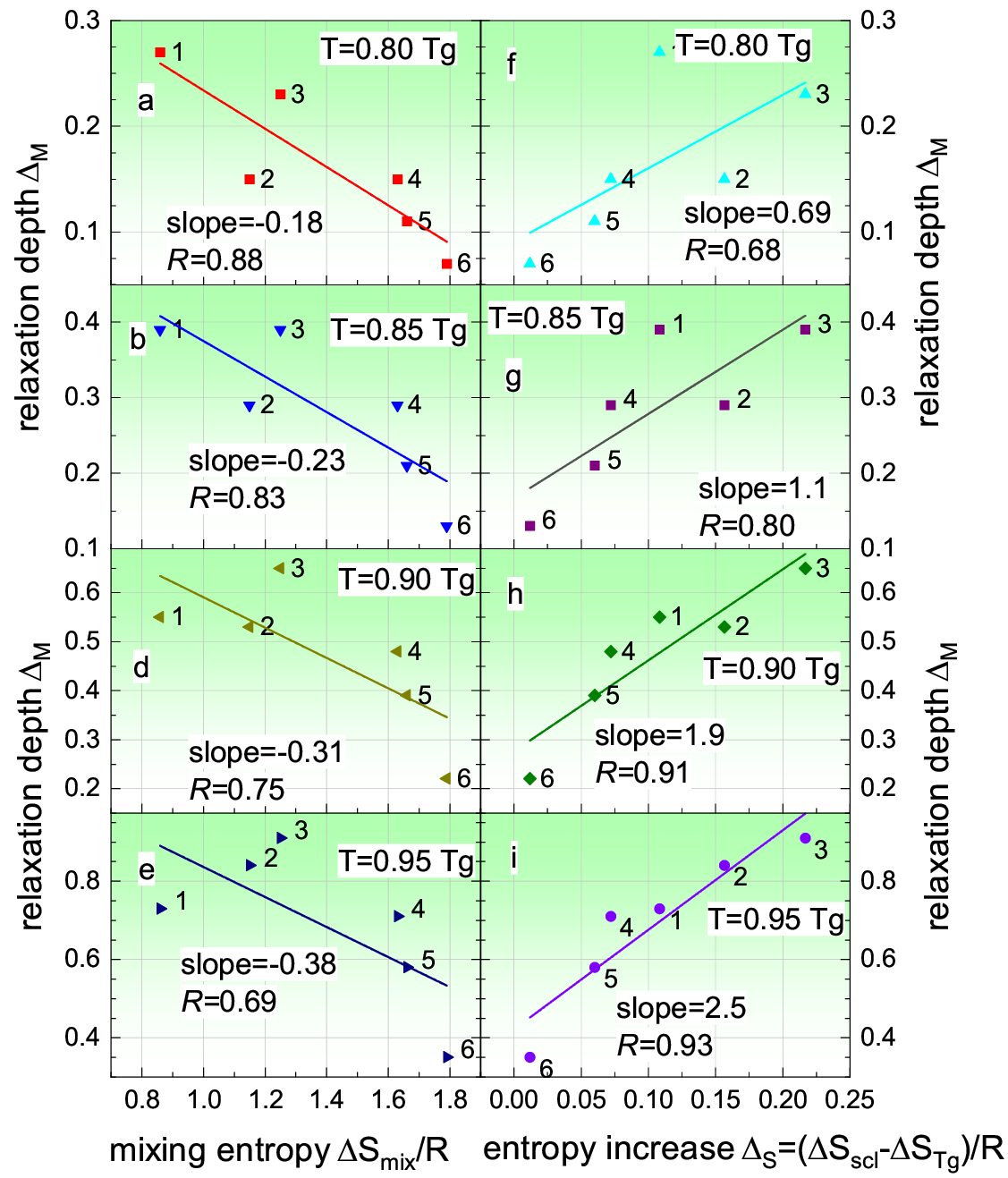}
\caption[*]{\label{Fig5.eps} Depth of stress relaxation $\Delta_M=1-M/M_0$ as a function of the mixing entropy $\Delta S_{mix}$ (panels a-e) and entropy rise in the SCL range $\Delta_S = (\Delta S_{scl}-\Delta S_{Tg})/R$ (panels f-i) at indicated temperatures of $0.80\;T_g$ (a,f),  $0.85\;T_g$ (b,g), $0.90\;T_g$ (d,h) and $0.95\;T_g$ (e,i). The lines give the least square fits. The corresponding slopes and Pearson's correlation coefficients $R$ are indicated.} 
\end{center}
\end{figure} 

The same trend is more clearly shown in panels (a)-(e) of Fig.\ref{Fig5.eps}. These panels show the stress relaxation depth $\Delta_M = 1 - M/M_0$ as a function of the mixing entropy $\Delta S_{mix}$, for different temperatures: $0.80T_g$, 0$.85T_g$, $0.90T_g,$ and $0.95T_g$. It is seen that in all cases $\Delta_M$ decreases with $\Delta S_{mix}$. The Pearson's  coefficients of $R\geq 0.73$ indicating quite reliable $\Delta_M(\Delta S_{mix})$-correlation. The slope of these dependences increases with normalized temperature.     

In principle, this trend is expected since it was found that an increase in the mixing entropy leads to a significant increase in the shear viscosity $\eta(T=T_g)$ at the glass transition temperature  \cite{MakarovIntermet2024} while the shear viscosity directly reflects the relaxation rate. Second, it was shown that an increase in the mixing entropy leads to a significant decrease of the excess entropy (both $\Delta S_{Tg}$  and $\Delta S_{scl}$, which are illustrated in Fig.\ref{Fig2.eps}b) calculated with  Eq.(\ref{Scalc}) \cite{Afonin2024}. Since the excess entropy directly reflects the disorder in glass, one would also expect that MGs with higher $\Delta S$ should be more prone to structural relaxation, which is consistent with the observations. Conversely, glasses with lower $\Delta S$ (i.e., with higher $\Delta S_{mix}$) should be structurally more ordered and more resistant to structural relaxation under thermal and other external impacts \cite{Afonin2024}. A number of peculiar characteristic of HEMGs, such as lower atomic mobility \cite{ChenIntermetallics2023}, sluggish diffusion \cite{DuanPRL2022,JiangNatComm2021} and slow crystallization kinetics \cite{JiangJA2016,YangMaterResLett2018}, as well as the slow dynamics of homogeneous flow \cite{ZhangScrMater2022}, are in line with this conclusion.

The high mixing entropy of many-component metal MGs indicates that the undercooled melt is highly relaxed due to the accommodative movements of different atoms into a large number of potential wells available in the  structure. This relaxed melt is then quenched to form a solid glass, which to a certain extent inherits the relaxed state of the melt. Thus, the higher the mixing entropy, the greater the degree of relaxation in the as-quenched glass should be.  
It is also worth noting that the mixing entropy does not depend on structural state of the glass and has the same value in both the amorphous and crystalline states.

On the other hand,  it is clear that the excess entropy of glass with respect to the counterpart crystal    plays an important role in the relaxation properties of MGs \cite{Afonin2024,MakarovIntermet2024,MakarovScrMater2024} and this should apply in full to their stress relaxation behavior.  To quantify this idea, we suggest a new dimensionless entropy-based parameter $\Delta_S = (\Delta S_{scl}-\Delta S_{Tg})/R$, where the entropies $\Delta S_{scl}$ and $\Delta S_{Tg}$ are defined in the same way as above (see Fig.\ref{Fig2.eps}b). The quantity $\Delta_S$ determines the rise of the entropy in the SCL range  upon heating from $T_g$ up to $T_x$ expressed in the $R$-units. If the glass starts to crystallize immediately after $T_g$ is reached, $\Delta_S$ is zero. The larger the increase in entropy in the SCL range, the greater $\Delta_S$ will be. A larger $\Delta_S$ indicates a greater degree of structural disorder and should, therefore, lead to a greater relaxation.
 
The entropies $\Delta S_{Tg}$ and $\Delta S_{scl}$ for the MGs under study determined from calorimetric measurements are listed in Table 1 along with the parameter $\Delta_S$, which was calculated as described above. This calculation is further used to plot the depth of stress relaxation  $\Delta_M$ as a function of the entropy parameter $\Delta_S$  at different normalized temperatures $T/T_g$ as shown in panels (h) to (i) of Fig.\ref{Fig5.eps}. It is seen that $\Delta_M$ increases with $\Delta_s$ in all cases. The Pearson's correlation coefficient $R$ is not very high for  $T/T_g=0.80\;T_g$ ($R=0.68$) but it is significantly larger, $R\approx 0.8$ to 0.9, for all other temperatures $0.85\;T_g\leq T\leq 0.95\;T_g$. Besides that, the slope $\frac{d\Delta_M}{d\Delta_S}$ substantially increases  from 0.69 at $T=0.80\;T_g$ to 2.5 at $T=0.80\;T_g$. 

Thus, the relaxation depth $\Delta_M$ does indeed increase significantly with the disorder parameter $\Delta_S$, and this relationship becomes stronger as the normalized temperature $T/T_g$ increases. In other words, the relaxation depth   increases with the amount of structural disorder, which is achieved in the SCL range. That is why the parameter $\Delta_S$ can be used for a rough estimate of stress relaxation behavior using calorimetric data. 

In the meantime, the stress relaxation data shown in Fig.\ref{Fig5.eps} were obtained for temperatures $T\leq 0.95\;T_g$, while the parameter $\Delta_S$ is calculated for the SCL range, that is, for \textit{higher} temperatures $T_g\leq T\leq T_x$ (see Fig.\ref{Fig2.eps}b). Therefore, it should be understood why $\Delta_S$ determines the relaxation behavior at temperatures below the glass transition temperature $T_g$. We suggest the following qualitative explanation. 

For this purpose, it should be noted that the excess entropy, as represented by either $\Delta S_{Tg}$ or $\Delta S_{scl}$, is directly proportional to the excess enthalpy $\Delta H$ of the glass with respect to the counterpart crystal \cite{Afonin2024}. The latter quantity, $\Delta H$, approximately equals the elastic energy of interstitial-type defects that are assumed to be "frozen-in" from the melt during glass production and  responsible for MGs' relaxation ability \cite{MakarovJETPLett2022,KobelevUFN2023}. Upon crystallization, these defects disappear and their elastic energy is dissipated into heat, which  equals the crystallization heat \cite{MakarovJETPLett2022,AfoninActaMater2016}. Thus, the  entropy parameter $\Delta_S$ should reflect  the  defect concentration in the SCL range. In this regard, it is worth noting that the degree of structural disorder, as indicated by the width of the X-ray structure factor, is indeed related to the concentration of defects \cite{MakarovIntermetallics2023}. Therefore, one can assume that the parameter $\Delta_S$ constitutes an indirect measure of the defect concentration frozen-in upon glass production. In this case, an increase of this concentration  should increase the depth of stress relaxation, as found in this investigation.

Overall, the relaxation depth $\Delta_M$ at a given temperature decreases with the mixing entropy $\Delta S_{mix}$ but increases with the entropy parameter of disorder $\Delta_S$.  In other words, the relaxation ability of MGs decreases with  $\Delta S_{mix}$ and increases with $\Delta_S$. These parameters play a key role in the relaxation behavior of MGs.  At that, the mixing entropy $\Delta S_{mix}$ characterizes the amount of disorder introduced during the preparation of the master alloy and remains unchanged during glass production. On the other hand, the disorder parameter $\Delta_S$ quantifies the structural state of a particular glass in the SCL state. 

It is interesting to note that the mixing entropy $\Delta S_{mix}/R$ is much larger than $\Delta_S$  and the ratio $\frac{\Delta S_{mix}}{\Delta_S}$ rapidly increases with $\Delta S_{mix}$ from $\approx 8$  the glass 1 to $\approx 149$ for the glass 6 (see Table 1). This fact underlines the importance of the mixing entropy in MGs' structural ordering and reflects its major role  in their relaxation  behavior. Since high-entropy MGs ($\Delta S_{mix}/R\geq 1.5$) display reduced relaxation ability, they can be considered promising for use in the  applications that require relatively high relaxation stability.

\section{Conclusions}

Calorimetric studies and measurements of torsional stress relaxation upon linear heating of six  metallic glasses (MGs) are carried out. The experiments were performed on conventional  and high entropy MGs, with a mixing entropy $\Delta S_{mix}$ covering a wide range from $0.86\;R$ to $1.76 \;R$, where $R$ is the universal gas constant. It is found that stress relaxation depth $\Delta_M$ decreases with increasing $\Delta S_{mix}$ at a given temperature normalized temperature $T/T_g$ ($T_g$ is the glass transition temperature). Consequently, high-entropy MGs with a mixing entropy of $\Delta S_{mix} > 1.5 R$ exhibit higher resistance to stress relaxation. At that, the rate of $\Delta_M$-decrease with $\Delta S_{mix}$ increases with $T/T_g$.

In order to characterize the reasons of this behavior, we introduced a new dimensionless entropy parameter of structural disorder $\Delta_S = (\Delta S_{scl}-\Delta S_{Tg})/R$, where $\Delta S_{Tg}$ is the excess entropy of glass with respect to the counterpart crystal calculated from calorimetric data at $T=T_g$ and $\Delta S_{scl}$ is the excess entropy corresponding to the end of the supercooled liquid (SCL) range, just below the crystallization onset. The parameter $\Delta_S$ thus defined characterizes the  increase of the excess entropy in the SCL range, which reflects an increase of  structural disorder upon heating in  this region. It shown that stress relaxation depth $\Delta_M$ at a given $T/T_g$ increases with $\Delta_S$ while the rate of $\Delta_M$-rise increases with $T/T_g$.  

It is concluded that the parameters $\Delta S_{mix}$ and $\Delta_S$ play a key role in the relaxation behavior of both conventional and high-entropy MGs, and may be used to predict their relaxation behavior. It is suggested that the disorder parameter $\Delta_S$  can be related to the concentration of defects frozen-in from the melt upon glass production.

\section*{Acknowledgements}

The work was supported by the Russian Science Foundation (project N\textsuperscript{\underline{o}}~23-12-00162). 



\end{document}